\documentclass[letterpaper]{IEEEtran}
\usepackage{graphicx,times, amsmath, amsfonts,comment}
\usepackage{amssymb,epstopdf}
\usepackage[noend]{algorithmic}
\usepackage{algorithm}
\usepackage{multirow}

\newcommand{\figref}[1]{Fig.~\ref{#1}}
\begin{document}

\title{Non-Orthogonal Multiple Access (NOMA) in Cellular Uplink and Downlink: Challenges and Enabling Techniques}
\author{Hina Tabassum,  Md Shipon Ali, Ekram Hossain, Md. Jahangir Hossain, and Dong In Kim
}

\maketitle

\begin{abstract}
By combining the concepts of superposition coding at the transmitter(s) and successive interference cancellation (SIC) at the receiver(s),
non-orthogonal multiple access (NOMA) has recently emerged as a promising multiple access technique for 5G wireless technology. In this article,  we first discuss the fundamentals of uplink and downlink NOMA transmissions and outline their key distinctions (in terms of  implementation complexity, detection and decoding  at the SIC receiver(s), incurred intra-cell and inter-cell interferences).
Later, for both downlink and uplink NOMA, we theoretically derive the NOMA dominant condition  for each individual user in a two-user NOMA cluster.
NOMA dominant condition refers to the condition under which the spectral efficiency gains of NOMA  are guaranteed compared to conventional orthogonal multiple access (OMA). The derived conditions provide direct insights on selecting appropriate users in two-user NOMA clusters.
The conditions are distinct for uplink and downlink as well as for each individual user.
Numerical results show the significance of the derived conditions  for the user selection in uplink/downlink NOMA clusters and provide a comparison to the random user selection. A  brief overview of the recent research investigations is then provided to highlight the existing research gaps. Finally, we discuss the potential  applications and key challenges of NOMA transmissions.

\end{abstract}


\section{Introduction}

Non-orthogonal multiple access (NOMA) is being considered as an enabling technique for 5G  cellular networks.
Until very recently, the state-of-the-art wireless communication systems have been utilizing  orthogonal multiple access (OMA) techniques, in which the resources are allocated orthogonally to multiple users. These techniques include, 
frequency-division multiple access (FDMA), time-division multiple access (TDMA), code-division multiple access (CDMA), and orthogonal frequency-division multiple access (OFDMA).  Ideally, in OMA, the intra-cell interference does not exist  due to the orthogonal resource allocation among
users. Also, for this reason, the  information of multiple users can be retrieved at a low complexity. Nonetheless, the number of served users is limited by the number of orthogonal resources. 

Conversely, NOMA\footnote{In this paper, NOMA refers to power-domain NOMA.} serves  multiple users  simultaneously using the same spectrum resources (i.e., radio channels); however, at the cost of  increased intra-cell interferences.
To mitigate the intra-cell interferences, NOMA exploits Successive Interference Cancellation (SIC) at the receivers~\cite{saito2013,ben2015}.  In addition, NOMA supports low transmission latency and signaling cost compared to  conventional OMA  where each user is obliged to send  a channel scheduling request to its serving base station (BS).   With these attractive features, NOMA is a potential access technology for 5G networks, where massive connectivity/coverage is required over limited radio resources.  

Over last couple of years, different bodies from industry and academia have proposed potential approaches to implement NOMA over existing LTE systems. As a part of {\em Mobile and wireless communications Enablers for the Twenty-twenty Information Society (METIS) project}, NTT DoCoMo proposed NOMA as a potential candidate for 5G radio access technology~\cite{saito2013}. 
A test-bed for a two-user downlink NOMA cluster was developed in \cite{ben2015}.
The experiments were performed by setting $5.4$ MHz bandwidth for NOMA users. The results were compared to those for two OMA users each having the transmission bandwidth of $2.7$ MHz. The results show the significance of NOMA over OMA in terms of aggregate as well as individual user's throughput.

In this article, we first describe the fundamentals of uplink and downlink NOMA transmissions and emphasize their key distinctions (in terms of  implementation complexity, detection and decoding  at the SIC receiver(s), incurred intra-cell and inter-cell interferences).  Then, for both  downlink and uplink NOMA,  we theoretically derive the NOMA dominant condition  for each individual user in a two-user NOMA cluster.
The NOMA dominant condition refers to the condition under which the spectral efficiency gains of NOMA  are guaranteed compared to OMA. 
The derived conditions are entirely distinct for uplink and downlink as well as for each individual user in a two-user NOMA cluster.
Numerical results show the significance of the derived conditions in selecting appropriate users for two-user uplink/downlink NOMA clusters and provide a comparison to the random user selection.
We then provide a  brief overview of the recent research investigations to highlight the existing research gaps.
Finally,  we highlight and discuss the potential  applications and key challenges of NOMA, such as multi-user grouping and power allocation, impact of  SIC error propagation, cooperative transmissions, and inter-cell interference management.  

\section{Working Principle and Key Distinctions of the Uplink and Downlink NOMA}
In this section, we detail the working principles and distinctions of both uplink and downlink NOMA transmissions.

\subsection{Working Principle of NOMA}
In general, NOMA allows the superposition of distinct message signals of users in a NOMA cluster. The desired message signal is then detected and decoded at the receiver (user in the downlink and BS in the uplink) by applying  SIC.

\subsubsection{Downlink NOMA}
In downlink,  the BS transmits the superimposed signal $x=\sum_{i=1}^{U} \sqrt{p_i} x_i$, where $x_i$ is the unit power message signal intended  for user $i$,  $p_i$ denotes the power allocated for user $i$, and $U$ denotes the total number of users in a NOMA cluster. The power allocated to a  user depends on the powers of other users due to the BS total power constraint, $P_t=\sum_{i=1}^{U} {p_i}$, where $P_t$ is the BS total power. The received signal at the $i$-th user is given by $y_i=h_i x +w_i$, where $h_i$ represents the channel gain between the BS and user $i$, and $w_i$ denotes the  Gaussian noise (with power spectral density $N_{0,i}$) at the  receiver for user $i$.

Unlike ``conventional water-filling", downlink NOMA utilizes a power allocation mechanism where high transmission powers are used for users with poor channel conditions and vice versa. Thus, at a given user in NOMA cluster, the strong interfering\footnote{Interference refers to the  intra-cluster interference within a NOMA cluster, unless stated otherwise.} signals are mainly due to the high power message signals of relatively weak channel users.  As such, to extract the desired signal, each user cancels the strong interferences by  successively decoding, remodulating,  and subtracting them from the received  signal $x$. Subsequently, the highest channel gain user cancels all  intra-cluster interferences, whereas the lowest channel gain user receives the interferences from all users within its cluster~\cite{saito2013,ben2015}.

\subsubsection{Uplink NOMA}
In uplink,  each user transmits its individual signal $x_i$ with a transmit power $p_i$ such that the received signal at the BS can be defined as $y=\sum_{i=1}^{U} \sqrt{p_i} h_i x_i +w$, where $w$ denotes the receiver noise (with power spectral density $N_0$) at the BS. The  power transmitted per user  is limited by the user's maximum battery  power. Different from downlink NOMA, all users can independently utilize their battery powers up to the maximum as long as the channel gains of the  users are sufficiently distinct. If the channel gains are too close, power control can  be used to boost up the performance of the user with better channel gain, while maintaining the performance of the users with weaker  channel gains at a certain level.

Note that, to apply SIC  and decode signals at the BS, it is crucial to maintain the distinctness of various message signals that are superposed within $y$. 
Since the channels of different users are different in the uplink, each message signal experiences distinct channel gain\footnote{The conventional uplink transmit power control (typically intended to equalize the received signal powers of users) may remove the channel distinctness and thus may not be feasible for uplink NOMA transmissions.}. 
As a result, the received signal power corresponding to the strongest channel user is likely the strongest at the BS. Therefore, this signal is decoded first at the BS  and experiences interference from all  users in the cluster with relatively weaker channels. That is,  the transmission of the highest channel gain user experiences  interference from all users within its cluster, whereas the transmission of  the lowest channel gain user receives zero interference from the users in its cluster. 

\begin{figure}[t]
\begin{center}
\includegraphics[width = 2.5in]{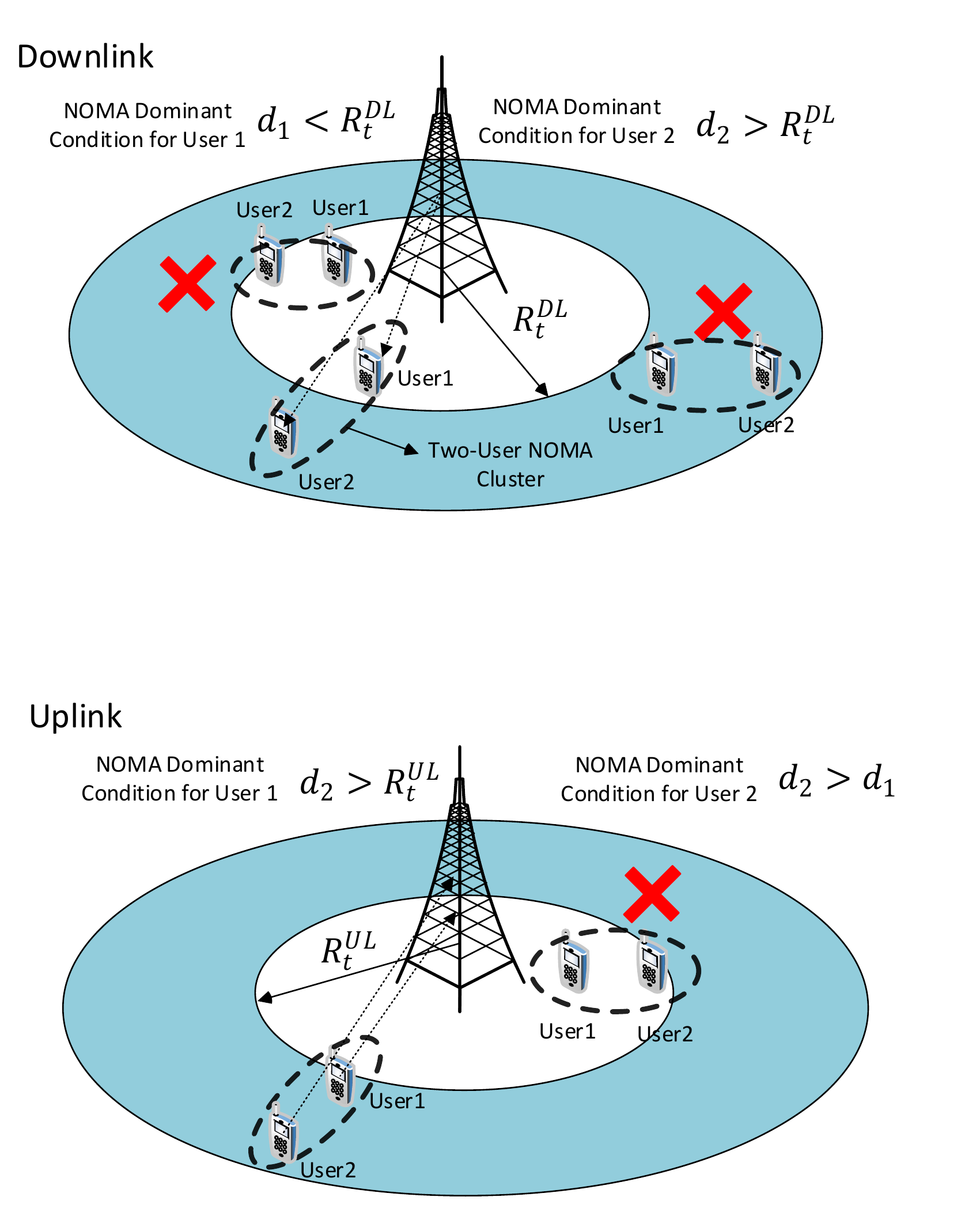}
\caption[c] {Graphical illustration of selecting users in a two-user downlink and uplink NOMA cluster such that the spectral efficiency of each individual user in NOMA outperforms OMA. $d_1$ and $d_2$ represent the distances of User~1 and User~2, respectively.}
\label{noma1}
\end{center}
\end{figure}

\subsection{Key Distinctions Between Uplink and Downlink NOMA}
\subsubsection{Implementation Complexity}
Downlink NOMA requires the implementation of sophisticated multi-user detection and interference cancellation schemes at the receiver of each user. This is a cumbersome task provided the  limited processing capability of users. However, in the uplink, it is relatively more convenient to implement multi-user detection and interference cancellation schemes on a  centralized entity (i.e., BS).

\subsubsection{Intra-cell/Intra-cluster Interference} 
The downlink intra-cell interference  at a user is experienced on  its own channel. Thus, a user with the strong downlink channel receives strong interference. In particular, the reason is two-fold: the strong channel (which is interfering) and relatively high transmit powers allocated for the messages of weak channel users. Similarly, a user with  weak downlink channel experiences low interference due to its own weak channel and the low transmit powers allocated for the messages of strong channel users. As such, the users with  {\em strong downlink channels are relatively more vulnerable} to intra-cell interference.  This is resolved by applying SIC at the users as will be explained later.

In the uplink, the BS receives transmissions from all users simultaneously. As such, the intra-cell interference to a user  is a function of the channel statistics of other users  within its cluster. Subsequently, the transmissions of {\em weak channel users  are relatively more vulnerable} to strong interferences that arise from the transmissions of strong channel users. This issue is however resolved by applying SIC at the BS as will be explained shortly.

\subsubsection{SIC at Receiver(s)}
In downlink NOMA, the strong channel users achieve throughput gains,  by successively decoding and canceling the messages of weak channel users, prior to decoding their desired signals. 
In the uplink, to enhance the throughput of weak channel users, the BS successively decodes and cancels the  messages of strong channel users, prior to decoding the signals of weak channel users. 


\subsubsection{Inter-cell interference}
Downlink multi-cell applications of NOMA will induce additional interference (from the neighboring co-channel BSs) at each individual user in a NOMA cluster. 
However, the  downlink inter-cell interference received at each individual user in NOMA is same as in OMA.
For illustration, consider a two-user NOMA cluster (say cluster~1) in which each individual user receives transmission from its serving  BS during time $T$.
Now assume a neighboring co-channel BS serves another two-user NOMA cluster with a total power $P$  during time $T$.  In such a case, the users in cluster~1 experience interference due to transmission power $P$ from the neighboring BS. In  an equivalent OMA system,  downlink transmission of power $P$  is allocated to each user for a time duration of $T/2$. Thus, the interference received at each individual user of cluster 1 in OMA is same as in NOMA; however, it may not be the same for all users in the cluster.



On the contrary, in uplink NOMA, a higher number of users in the NOMA clusters served by neighboring  co-channel BSs result in a higher amount of interference at the BS of interest. Consequently, the uplink multi-cell interference in NOMA is directly proportional to the number and the transmit powers of users per NOMA cluster of neigboring co-channel BSs, and is therefore, different from OMA.

\section{User Selection: Uplink vs. Downlink NOMA}
In this section,  for both uplink and downlink, we derive a criterion for the selection of appropriate users in two-user NOMA clusters.
The criterion guarantees the individual throughput gains of each user in the cluster, compared to OMA.   Numerical results  are presented to show the significance of the proposed user selection criterion over random user selection in two-user uplink and downlink NOMA clusters.

\subsection{Downlink User Selection}
\subsubsection{System Set-up}
We consider two users  located at $d_1$ and $d_2$ such that $d_1<d_2$ with their average channel gains  given by $d_1^{-\alpha}$ and $d_2^{-\alpha}$, respectively. Note that the user at $d_1$ can be considered as the {\em strong user} and the user at $d_2$ can be considered as the {\em weak user}, respectively. Let $P$ denote the  normalized transmit power with respect to the noise power spectral density at the receiver (i.e., $P = P_t/N_0$) and $\alpha$ denote the path-loss exponent. The users located at $d_1$ and $d_2$ are referred to as $U_1$ and $U_2$, respectively. 
In NOMA, the spectral efficiency of transmission to $U_1$ and $U_2$ for a time duration $T$ can be computed, respectively, as follows:
\begin{align}\label{noma}
&C^{(\mathrm{noma})}_1 =\mathrm{log}_2(1+{a_1 P d_1^{-\alpha}})
\nonumber\\&
C^{(\mathrm{noma})}_2 =\mathrm{log}_2\left(1+\frac{a_2 P d_2^{-\alpha}}{a_1 P d_2^{-\alpha}+1}\right)
\end{align}
where $a_1$ and $a_2$ shows the portion of the BS transmit power allocated for $U_1$ and $U_2$, respectively, and $a_1+a_2=1$. Without loss of generality, the duration $T$ is taken as unity.

For comparison, we consider a TDMA-based OMA system, where each user receives its downlink transmission for a duration of $T/2$. The spectral efficiency of each user can thus be calculated as follows\footnote{Note that the spectral efficiency in TDMA could be optimized with respect to the time allocation.  That is, we may allocate more time to user 1,  while less time to user 2. However, for simplicity, here we assume equal time allocation for the two users.}:
\begin{equation}\label{oma}
C^{(\mathrm{oma})}_j =0.5~\mathrm{log}_2(1+ P d_j^{-\alpha}),\:\:\:\:\forall j=1,2.
\end{equation}


\begin{figure*}
\begin{minipage}{0.5\textwidth}
\includegraphics[width = 3.5in]{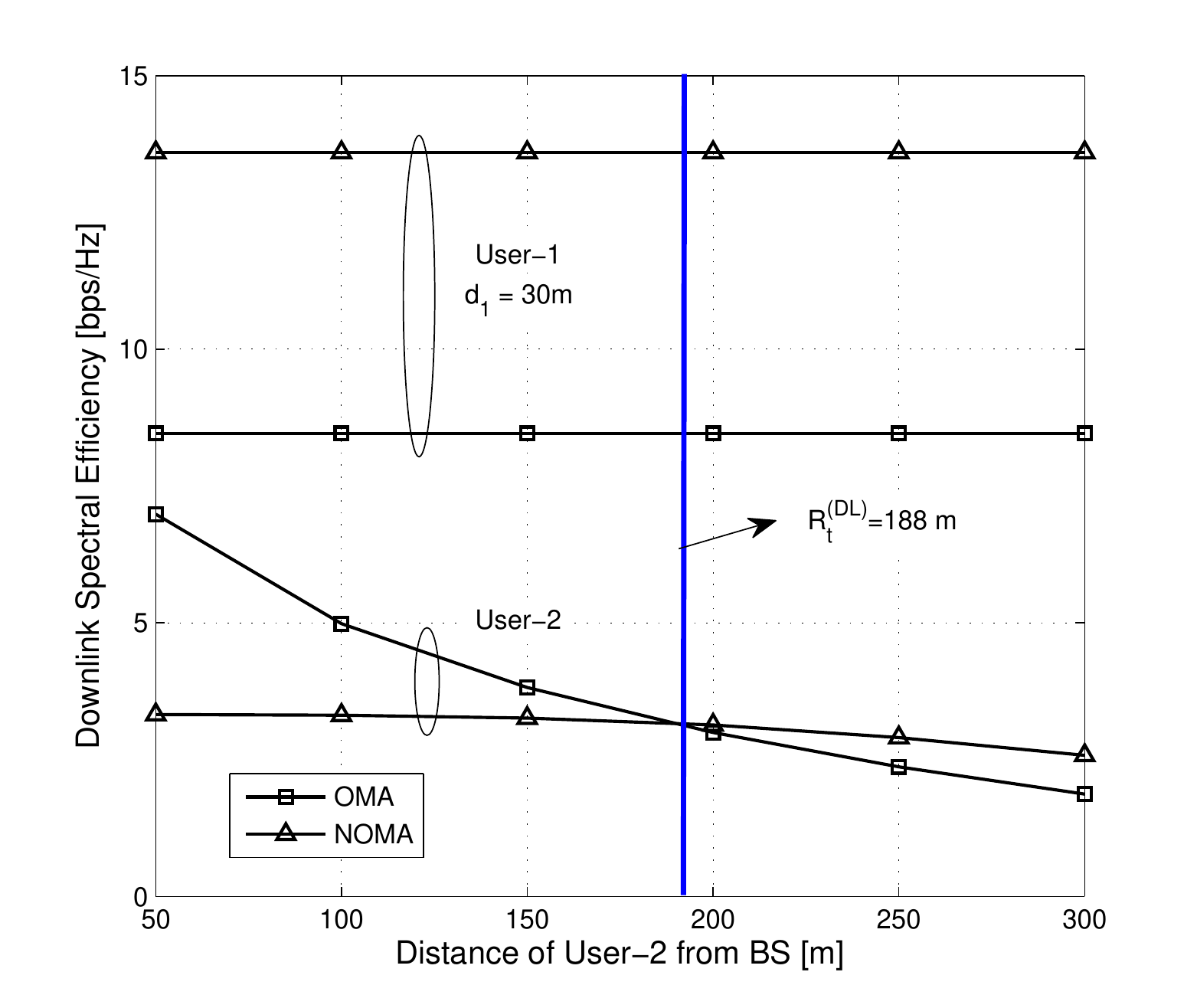}
\end{minipage}
\hfill
\begin{minipage}{0.5\textwidth}
\includegraphics[width = 3.5in]{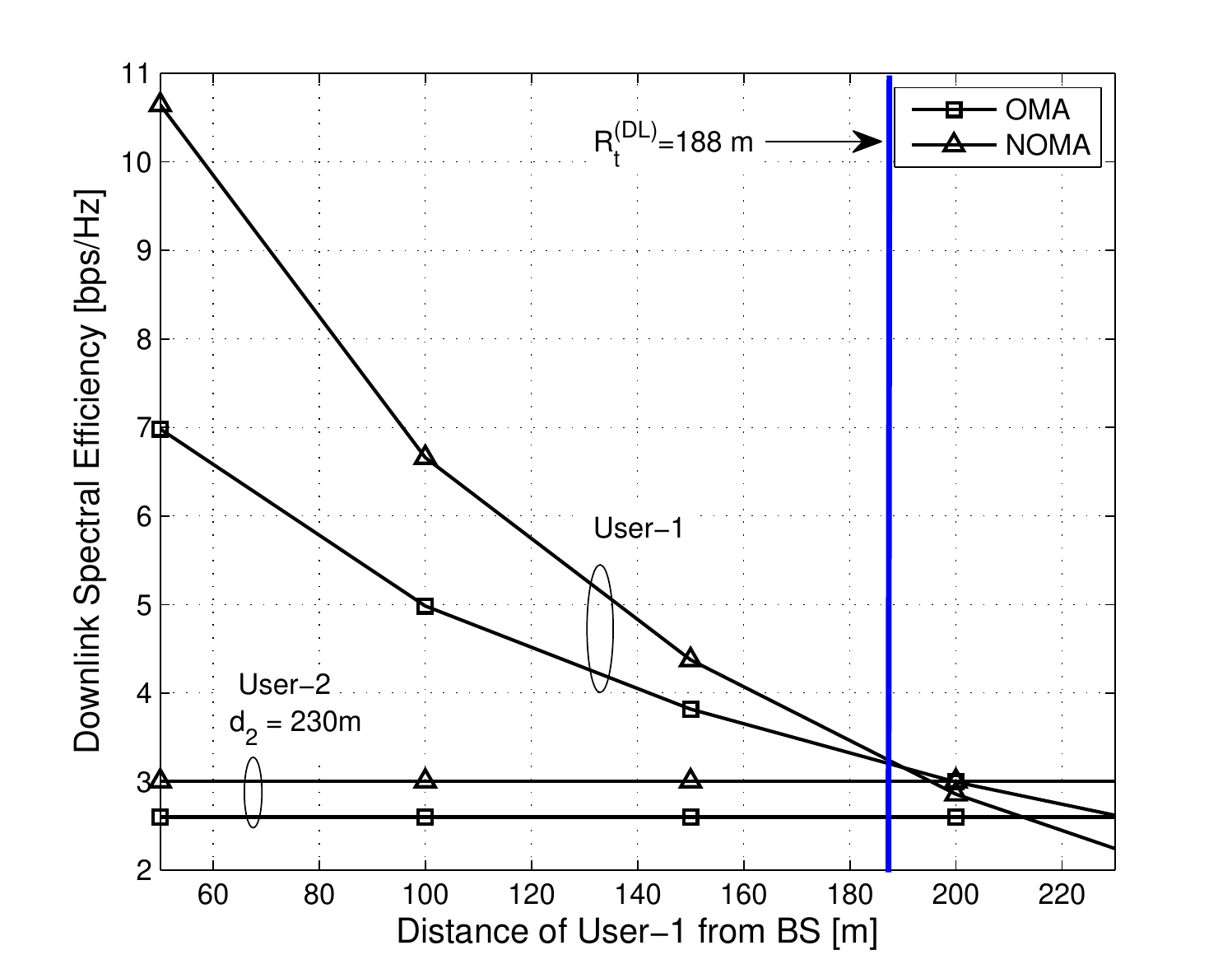}
\end{minipage}
\caption[c] { Downlink spectral efficiency of transmission to individual users as a function of (a)~distance of $U_2$ from the BS (for $d_1=30$~m, $a_1=0.1$, $P_t=10$~W, $N_0=10^{-10}$~W/Hz, $P=P_t/N_0$, $\alpha=4$), (b)~distance $U_1$ from the BS (for $d_2=230$~m, $a_1=0.1$, $P_t=10$~W, $N_0=10^{-10}$~W/Hz, $P=P_t/N_0$, $\alpha=4$).}
\label{nomadl}
\end{figure*}

\subsubsection{User Selection Criterion}
In this subsection, we derive a minimum distance $R_t^{(\mathrm{DL})}$ (see \figref{noma1}) beyond which the NOMA throughput gains are guaranteed for a user with weak channel gain (i.e., $U_2$, as shown in \figref{noma1}). Further, we note that the user with  strong channel gain (i.e., $U_1$, as shown in \figref{noma1}) should  remain within $R_t^{(\mathrm{DL})}$ to achieve the throughput gains of NOMA.
Subsequently, the two-user NOMA pair should be formed such that the strong and weak channel users are located within  $R_t^{(\mathrm{DL})}$ and beyond $R_t^{(\mathrm{DL})}$, respectively.

In order to find the boundary distance at which NOMA outperforms OMA for each individual user in the cluster, we utilize the condition $C^{(\mathrm{noma})}_j > C^{(\mathrm{oma})}_j, \:\:\forall j$ and substitute the values from \eqref{oma} and \eqref{noma}. After some algebraic manipulations, we note that 
$d_1 \leq R_t^{(\mathrm{DL})}$ and $d_2 > R_t^{(\mathrm{DL})}$ are necessary conditions to guarantee the gains of NOMA over OMA for each individual user.  The value of $R_t^{(\mathrm{DL})}$ can be derived as follows:
\begin{align}\label{r1}
R_t^{(\mathrm{DL})}=
\left(\frac{1-2a_1}{P a_1^2}\right)^{-\frac{1}{\alpha}}.
\end{align}
From \eqref{r1}, we can see that $R_t^{\mathrm{DL}} >0$ {\bf iff} $a_1<0.5$, i.e, the NOMA gains can be  achieved {\bf iff} we allocate low powers to strong channel users. Further, as $a_1$ increases, $R_t^{(\mathrm{DL})}$ increases due to the squaring effect of $a_1$ in \eqref{r1}. {\em 
$R_t^{\mathrm{DL}}$ also depicts the minimum coverage area that a BS  should be capable of serving. In particular, small cell BSs with reduced coverage areas (that are less than $R_t^{\mathrm{DL}}$) may not be suitable for downlink  NOMA transmissions.}

%

\begin{figure*}
\begin{minipage}{0.5\textwidth}
\includegraphics[width = 3.5in]{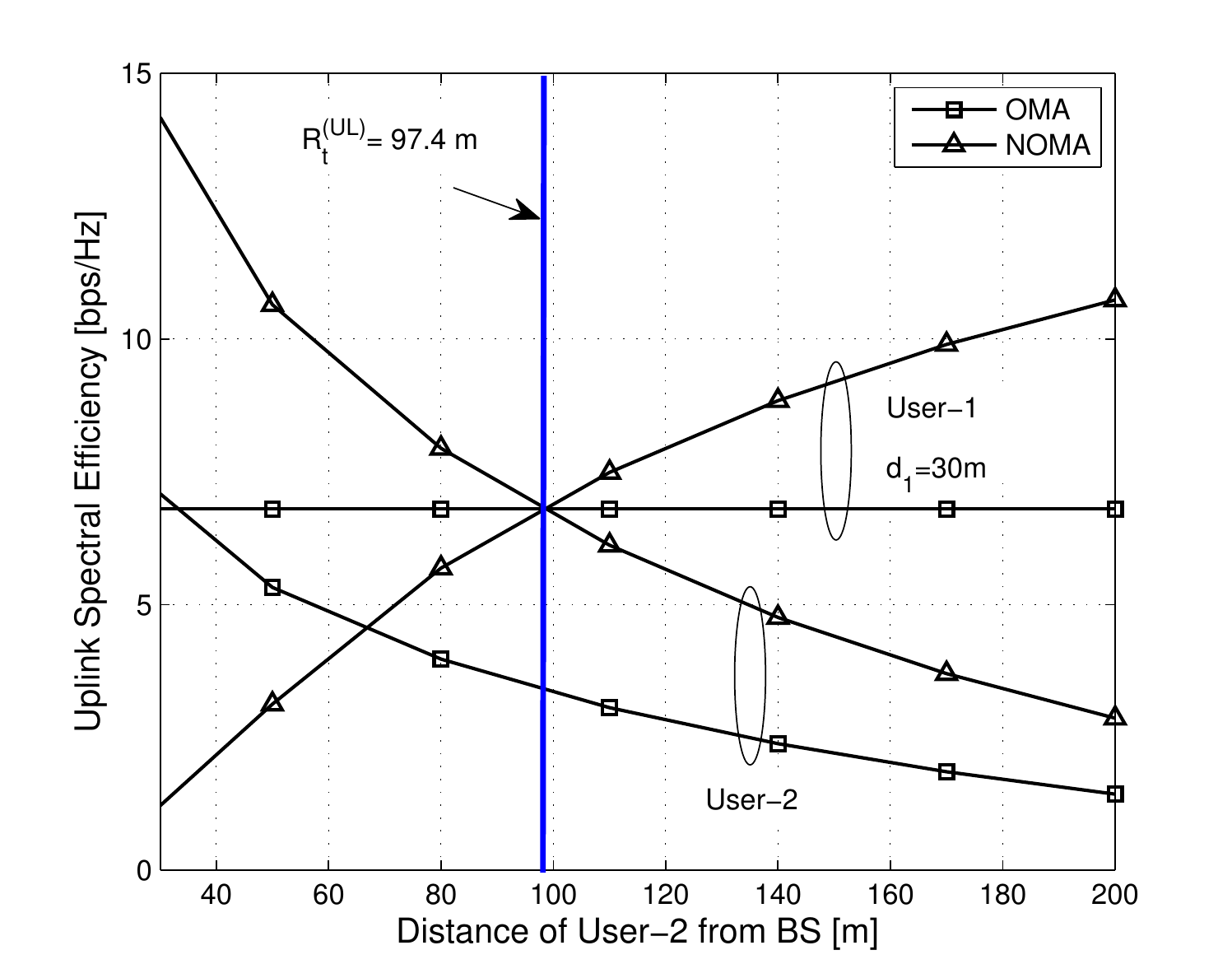}
\end{minipage}
\hfill
\begin{minipage}{0.5\textwidth}
\includegraphics[width = 3.5in]{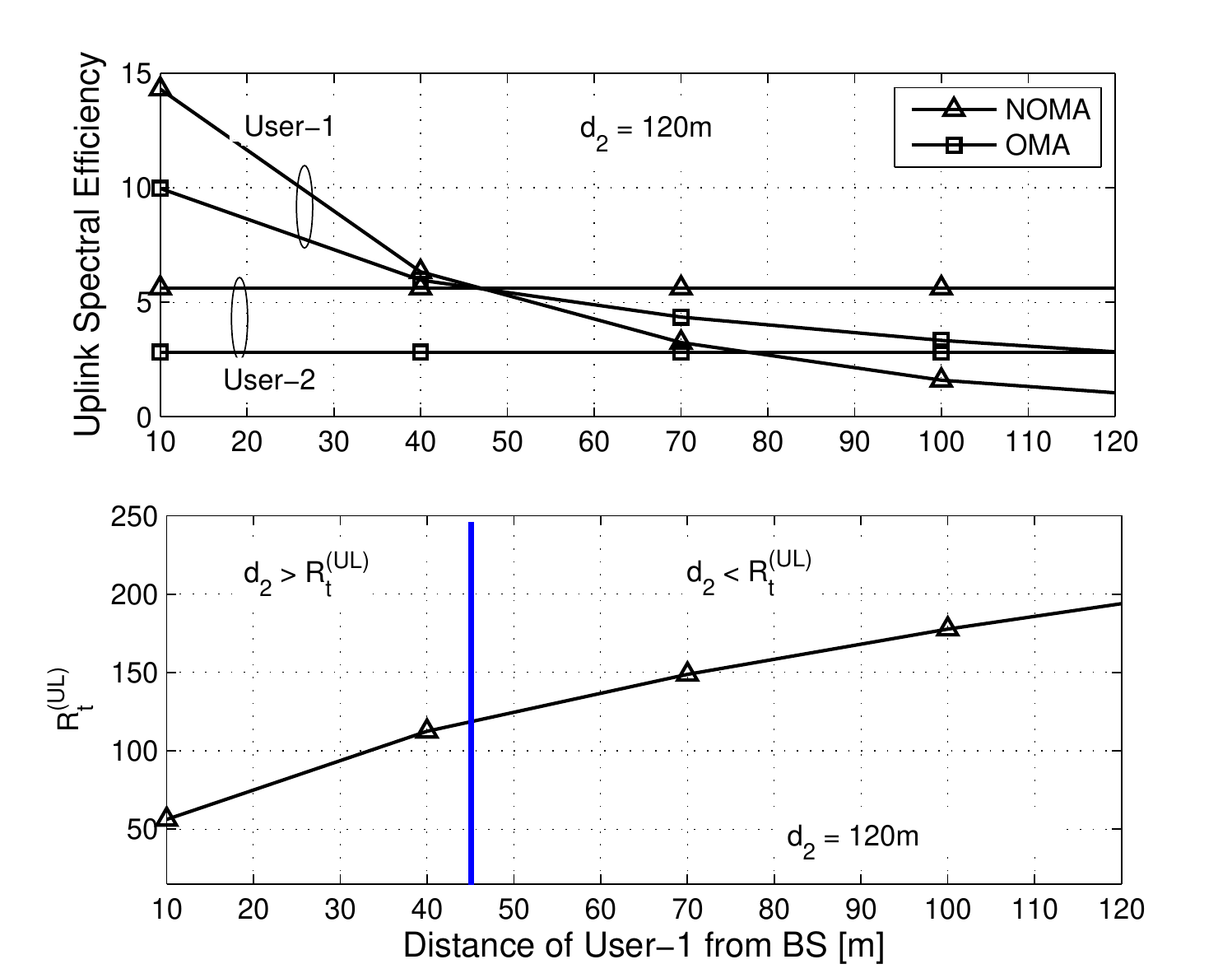}
\end{minipage}
\caption[c] { Uplink spectral efficiency of transmission to each individual user in a NOMA cluster as a function of (a)~distance of $U_2$ from the BS (for  $P_t=1$~W, $N_0=10^{-10}$~W/Hz, $P=P_t/N_0$, $\alpha=4$, $d_1$ = 30~m), (b)~distance of $U_1$ from the BS (for  $P_t=1$~W, $N_0=10^{-10}$~W/Hz, $P=P_t/N_0$, $\alpha=4$, $d_2$ = 120~m).}
\label{nomaul}
\end{figure*}

\subsubsection{Numerical Results}
\figref{nomadl}(a) demonstrates the spectral efficiency of  $U_1$ and $U_2$ as a function  of $d_2$ when $d_1=30~$m is fixed.  As $d_2$ increases, the throughput of $U_2$ reduces due to path-loss degradation, in both OMA and NOMA. However, for $U_2$, the gains of NOMA become evident only when $d_2 \geq R^{(\mathrm{DL})}_{t}=188$~m. This confirms the accuracy of the derived theoretical result in \eqref{r1}. On the other hand, the spectral efficiency of  $U_1$ is independent of $d_2$ and thus remains unchanged. Note that, $d_1 < R^{(\mathrm{DL})}_{t}$; therefore, NOMA outperforms OMA for $U_1$.
Similarly, \figref{nomadl}(b) captures the spectral efficiency of  $U_1$ and $U_2$ as a function  of $d_1$ when $d_2=230$~m is fixed.  As $d_1$ increases, the throughput of $U_1$ reduces due to path-loss degradation, in both OMA and NOMA. However, for $U_1$, the gains of NOMA become evident only when $d_1 \leq R^{(\mathrm{DL})}_{t}=188$~m.  The spectral efficiency of  $U_2$ is independent of $d_1$ and thus remains unchanged. Note that, $d_2 > R^{(\mathrm{DL})}_{t}$; therefore, NOMA outperforms OMA for $U_2$.
 
\subsection{Uplink User Selection}
\subsubsection{System Set-up}
We consider two users  located at $d_1$ and $d_2$ such that $d_1<d_2$. Let $P_1$ and $P_2$ denote the transmit powers of $U_1$ and $U_2$ normalized by their receiver noise powers, respectively.  In conventional OMA, the spectral efficiency of $U_1$ and $U_2$ can be calculated as in \eqref{oma} by replacing $P$ with $P_j, \:\:\forall j=1,2$. In NOMA, the  spectral efficiency of $U_1$ and $U_2$ can be computed, respectively, as $C^{(\mathrm{noma})}_1 =\mathrm{log}_2\left(1+\frac{P_1 d_1^{-\alpha}}{P_2 d_2^{-\alpha}+1}\right)$ and $C^{(\mathrm{noma})}_2 =\mathrm{log}_2(1+{ P_2 d_2^{-\alpha}})$.

\subsubsection{User Selection Criterion}
In uplink, first we note that NOMA always outperforms OMA for the weak channel user $U_2$ given that there is no intra-cell interference. However, the throughput gains of NOMA for a user with strong channel $U_1$ can be guaranteed {\bf iff} $U_2$ is located beyond $R_t^{(\mathrm{UL})}$.
As such, we derive $R_t^{(\mathrm{UL})}$ which allows appropriate selection of $U_2$ given the distance of $U_1$.

To find the boundary distance $R_t^{(\mathrm{UL})}$, we utilize the condition $C^{(\mathrm{noma})}_j > C^{(\mathrm{oma})}_j,\:\:\forall j$. After some algebraic manipulations, it can be seen that NOMA is always better than OMA for $U_2$. This is different from downlink. Further, 
the distance of $U_2$ should be at least $R^{(\mathrm{UL})}_t$  to guarantee the NOMA throughput gains of $U_1$. The value of $R^{(\mathrm{UL})}_t$ is derived as follows:
\begin{align}\label{r2}
R^{(\mathrm{UL})}_t=
\left(\frac{\sqrt{1+P_1 d_1^{-\alpha}}}{P_2}\right)^{-\frac{1}{\alpha}}.
\end{align}
Note that, the result in \eqref{r2} is different from that for downlink. The reason is that the distance of $U_2$ directly impacts the performance of $U_1$ in the uplink NOMA cluster. 

\subsubsection{Numerical Results}
\figref{nomaul}(a) demonstrates the spectral efficiency of  $U_1$ and $U_2$ as a function  of $d_2$ when $d_1=30$m is fixed.  
For ease of exposition and without loss of generality, we consider $P_1=P_2$. Contrary to OMA, we can see that the throughput  of  $U_1$ is a function of $d_2$ in NOMA. Since the BS decodes the strongest user first in NOMA, the transmission of $U_2$ creates interference to $U_1$. Therefore, for $U_1$, the gains of NOMA become evident only when $d_2 \geq R^{(\mathrm{UL})}_{t}$. This also confirms the accuracy of the derived theoretical result in \eqref{r2}.
Further, as $d_2$ increases, the throughput of $U_2$ reduces due to path-loss degradation in both OMA and NOMA. It can be observed that  NOMA always outperforms OMA for  $U_2$.

\figref{nomaul}(b) shows the spectral efficiency of  $U_1$ and $U_2$ as a function  of $d_1$ when $d_2=120$~m is fixed.  The spectral efficiency of  $U_2$ is independent of $d_1$ and thus remains unchanged. Further, NOMA always outperforms OMA for $U_2$. However, an interesting observation is that $R^{(\mathrm{UL})}_{t}$ increases with increasing $d_1$. That is, if $d_1$ increases, $R_t^{(\mathrm{UL})}$  increases which implies that $d_2$ should be increased beyond $R_t^{(\mathrm{UL})}$. Since in this figure $d_2$ is fixed, the NOMA gains of $U_1$ can be observed only when the condition $d_{2} > R^{(\mathrm{DL})}_{t}$ remains satisfied (see top figure). The  region in which the condition is valid is marked clearly in the bottom figure.

\subsection{Summary}
\begin{figure*}
\begin{minipage}{0.5\textwidth}
\includegraphics[width = 3.5in]{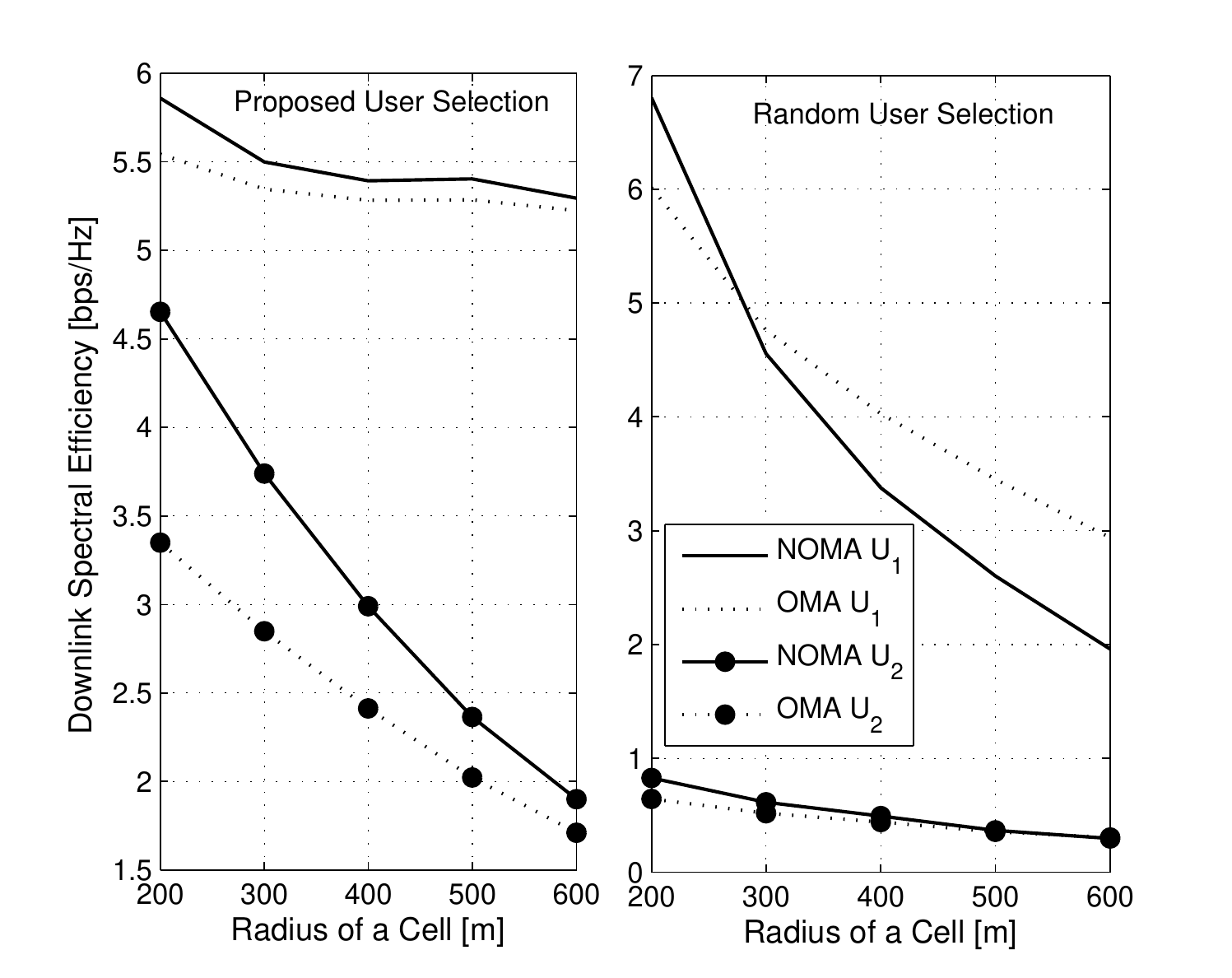}
\end{minipage}
\hfill
\begin{minipage}{0.5\textwidth}
\includegraphics[width = 3.5in]{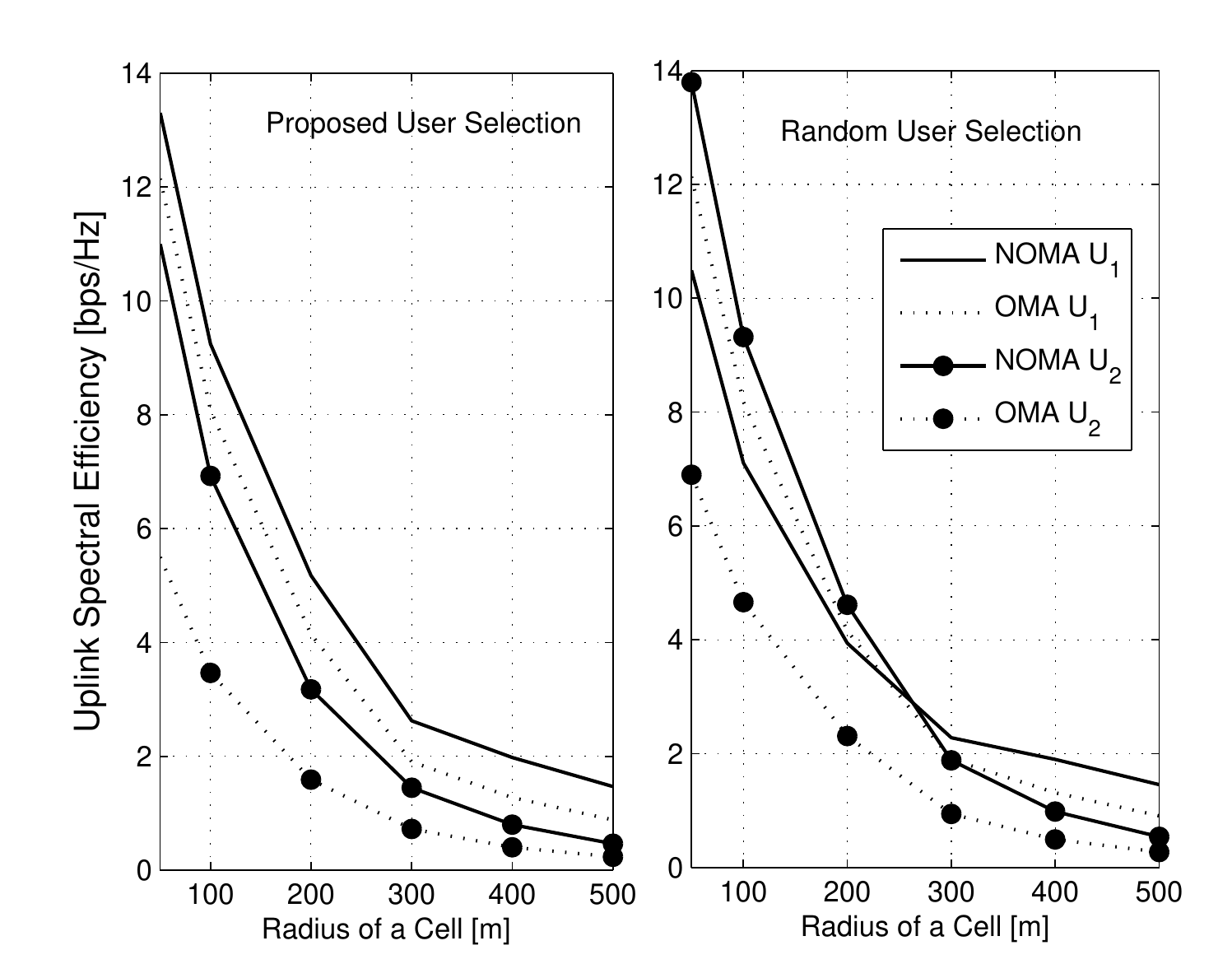}
\end{minipage}
\caption[c] {Comparison of random versus proposed user selection in (a)~downlink NOMA  (for  $a_1=0.1$, $P_t=10$~W, $N_0=10^{-10}$~W/Hz, $P=P_t/N_0$, $\alpha=4$), (b)~uplink NOMA (for  $P_t=1$~W, $N_0=10^{-10}$~W/Hz, $P=P_t/N_0$, $\alpha=4$).}
\label{proposed}
\end{figure*}

It can be concluded that, for a fixed power allocation,  the selection of users in a two-user downlink NOMA cluster should follow the conditions ($d_2 > R^{(\mathrm{DL})}_{t}$, $d_1 < R^{(\mathrm{DL})}_{t}$, and $d_1 < d_2$), to guarantee the performance gains of NOMA over OMA for each  user.
Similarly, the selection of users in a two-user uplink NOMA cluster should follow the conditions ($d_2 > R^{(\mathrm{UL})}_{t}$ and $d_1 < d_2$), to guarantee the performance gains of each  user. In this context, \figref{proposed} provides a comparison of the proposed  and the random user selection in uplink and downlink two-user NOMA clusters. It can be seen that the random user selection cannot guarantee the NOMA performance gains of $U_1$ in uplink and downlink, compared to OMA. However, the proposed user selection guarantees the performance gains of NOMA for each individual user in the cluster.

\section{State-of-the-Art: Uplink and Downlink NOMA}

In this section, we provide an overview of the recent research studies that dealt  with uplink and downlink NOMA transmissions. We review and categorize these studies mainly based on their power allocation methods, user clustering/pairing methods, and applicability of MIMO transmissions (Table~I). 

\subsection{State-of-the-Art: Downlink NOMA}
The concept of NOMA  was initially proposed in \cite{saito2013} for downlink transmissions. Various practical aspects, such as multi-user scheduling, impact of error propagation in SIC, overall system overhead, and user mobility were discussed. System level simulations were conducted in \cite{ben2015} to highlight the benefits of two-user NOMA over  OMA, in terms of overall system throughput as well as individual user's throughput. 
Closed-form expressions for the probability of a user  to be in NOMA cluster, ergodic sum-rate and outage probability   were presented in \cite{ding2014} considering static power allocations. It was concluded that without proper selection of target data rate, a NOMA user may always experience outage.

In \cite{ding2015}, the throughput gains of cooperative NOMA (in which the strong channel user relays the information of weak channel user) were investigated. It was shown that grouping users with distinct channel gains in a NOMA cluster provide significant performance gains over OMA.
Another interesting study is \cite{liu} where   cooperative  NOMA was applied to wireless-powered systems. Based on user's distances, grouping of users is  performed first. Then, three  user  selection  schemes  were investigated, i.e., (i)~pairing of the nearest users from each group, (ii) pairing of nearest user from one group and the farthest  user from another group, and (iii) arbitrary user pairing. The direct link was used to transfer energy from the BS. A cooperative data link was established for the lower channel gain user via the higher channel gain users. Closed-form expressions for  the  outage  probability  and  system
throughput for a two-user NOMA cluster were derived.

In \cite{dingearly}, the user pairing was investigated  considering  fixed power allocations (F-NOMA) and cognitive radio inspired power allocations (CR-NOMA). In F-NOMA, any two users could make a NOMA pair  based on their channel gains.  While in CR-NOMA, a weak channel user opportunistically gets paired with the strong channel user provided  that the interference caused by the strong user will not harm the quality-of-service (QoS) requirement of the weak channel user. It was observed that  CR-NOMA prefers to pair a strongest user with the second strongest user, whereas F-NOMA pairs a strongest user with the weakest user in the system.

Interestingly, the application of NOMA in MIMO systems boosts the performance gains of NOMA even further. In the sequel, a recent study  is \cite{dingmimo}, where the outage probability and throughput performance gains of MIMO-NOMA over MIMO-OMA systems were investigated. Another interesting study is \cite{qi} where an optimal as well as a sub-optimal power allocation solution was proposed to maximize the ergodic sum throughput of a two-user MIMO-NOMA system.


\subsection{State-of-the-Art: Uplink NOMA}

A general concept of uplink NOMA was presented in~\cite{endo2012} by considering transmission power control at users while utilizing the minimum mean squared error (MMSE)-based SIC decoding at the BSs. The authors evaluated the sum throughput and individual users' throughputs in a two-user NOMA cluster. 
Later, subcarrier and power allocation problem was studied in \cite{imari2014} to maximize the system throughput considering maximum transmit power  and minimum rate constraints. Due to the non-convex nature of the problem, a single user water-filling (SUWF)-based sub-optimal solution was proposed. In \cite{zhang2016}, an uplink power back-off policy was proposed to distinguish users in a NOMA cluster with nearly similar signal strengths (given that traditional uplink power control is applied). Closed-form analysis was performed for ergodic sum-rate and outage probability of  two users in a  NOMA cluster.

A sub-band based scheduling algorithm with single carrier (SC)-FDMA was proposed for a two-user NOMA cluster in \cite{li2015}. The power control was performed in two steps: (i)  traditional transmission power control (TPC) depending on the number of the scheduled sub-bands and (ii) static power  to diversify the received powers of different NOMA users. In \cite{chen2014}, an enhanced proportional fair  algorithm was proposed for user scheduling. System level simulations were conducted to show the gains of the proposed algorithm in multi-cell scenarios. To increase the fairness among users,  the concept of fractional frequency reuse was  exploited.

\subsection{Summary}
To date, most of the research investigations  have been conducted either for downlink or for uplink scenario considering a small-scale system with perfect SIC at the receivers. Further, the problem of user pairing and power control have been dealt separately in various set-ups (as detailed in Table~I). However,
there are no comprehensive studies where the problem of joint user clustering and power allocation is treated optimally. In this direction, a very recent and relevant study is  \cite{shipon} where we have proposed a sub-optimal user-clustering scheme and derived optimal power allocations for  multi-user uplink/downlink NOMA systems.  Another interesting study is \cite{higuchi} where the performance of a MIMO-NOMA system was evaluated in  both uplink and downlink.  To eliminate the inter-beam interference, spatial filtering was employed prior to the SIC processing.  Consequently, a framework to speculate the impact of optimal user clustering and power allocations in MIMO-NOMA system would be of immediate interest.



\begin{table*}[]
\centering
\caption{Summary of state-of-the-art of uplink and downlink NOMA}
\label{my-label}
\begin{tabular}{|c|c|c|c|c|c|c|}
\hline
Ref.                 & Main contributions                                                                                                & \begin{tabular}[c]{@{}c@{}}Transmission \\ scenario\end{tabular} & \begin{tabular}[c]{@{}c@{}}Power \\ allocation\end{tabular}      & User grouping                                                                          & \begin{tabular}[c]{@{}c@{}}Max. users in\\ NOMA cluster\end{tabular}       & 
\begin{tabular}[c]{@{}c@{}}Number of\\ antennas\end{tabular}                                                             \\ \hline \hline                                                               
$\cite{ding2014}$    & \begin{tabular}[c]{@{}c@{}} Performance Analysis \end{tabular}                                   
& Downlink                                                         
& Static                                                             
& $\times$                                                                               & Any                                                                      
& Single                                                                       \\ 
\hline
$\cite{liu}$         
& \begin{tabular}[c]{@{}c@{}}Cooperative NOMA \\ in SWIPT\end{tabular}                             & Downlink                                                         
& Static                                                            
& Sub-optimal                
& Two                                                                        
& Single    
\\ \hline
$\cite{dingmimo}$    
& Performance Analysis                                     
& \begin{tabular}[c]{@{}c@{}}Downlink\end{tabular}    
& Static                                                          
& $\times$                                                                               & Two                                                                      
& MIMO                                                                                                                                           \\ \hline
$\cite{qi}$          
& MIMO-NOMA                                            
& Downlink                                                         
& \begin{tabular}[c]{@{}c@{}}Dynamic \\ (Optimal)\end{tabular} 
& $\times$                                                                               & Two                                                                       
& MIMO                                                                                                                                      \\ \hline
$\cite{imari2014}$   & \begin{tabular}[c]{@{}c@{}} Joint Channel\\ and Power Allocation\end{tabular}                                                  
& Uplink                                                           
& Dynamic
& $\times$                                                                               & Any
& Single                                                                       \\ \hline
$\cite{zhang2016}$   
& Performance Analysis                                    
& Uplink                                                           
& Dynamic                                                           
& $\times$                                                                               & Two                                                                       
& Single                                                                      \\ \hline
$\cite{li2015}$      
& \begin{tabular}[c]{@{}c@{}}Sub-band Scheduling \\ in SC-FDMA\end{tabular}                                                                                              & Uplink                                                           
& Static                                                            
& Sub-optimal
& Two                                                                        
& Single                                                                      
                                                           \\ \hline
$\cite{shipon}$      
& \begin{tabular}[c]{@{}c@{}} Joint User Clustering \\and Power Allocation
\end{tabular} 
& \begin{tabular}[c]{@{}c@{}}Downlink\\ and uplink\end{tabular}    
& Optimal    
& Sub-optimal 
& Any  
& Single      
\\\hline
$\cite{higuchi}$    
& MIMO-NOMA                                            
& \begin{tabular}[c]{@{}c@{}}Downlink\\ and uplink\end{tabular}    
& Static                                                           
& $\times$                                                                               & Any                                                                      & \begin{tabular}[c]{@{}c@{}}MIMO in downlink,\\ and MISO in uplink\end{tabular}
\\\hline 
\end{tabular}
\end{table*}

\section{Existing Challenges and Future Research Directions}

In this section,  we highlight fundamental as well as foreseen challenges in the implementation of uplink and downlink NOMA. These include SIC propagation error, multi-user clustering, power allocation, and opportunistic NOMA in multi-cell scenarios. Also, in the same context, we accentuate potential applications of NOMA to overcome these challenges. 

\subsection{SIC Propagation Error} 
The performance of NOMA relies mainly on the successful SIC of the  strong interfering signals.  However, if any of these signals is detected erroneously, the  remaining signals may also be detected erroneously. As such, the intra-cell interference may not be eliminated appropriately and the desired signal of the receiver may not be decoded. It is thus of prime relevance to utilize more stringent performance metrics such as rate coverage/outage where the impact of these imperfections/decoding failures can be captured in a more accurate manner. 
To overcome  the drawbacks of SIC-based decoding, new interference cancellation schemes are also of special interest. 
Further, it is plausible to adapt the size of NOMA clusters as a function of the  incurred SIC error.  

\subsection{Multi-Cell NOMA Transmissions}
 NOMA is well-known to outperform OMA in a single-cell set-up. However, the validity of this conclusion is unknown for  multi-cell scenarios.   For instance, in the downlink multi-cell NOMA,  cell-edge users are more vulnerable to the transmissions of  neighboring BSs rather than the intra-cell interference. This is in sharp contrast to the single cell NOMA where the cell-edge users experience only intra-cell interference. Further, in the uplink multi-cell NOMA, inter-cell interference received at the given BS  is directly proportional to the number of users per NOMA cluster in neighboring cells.  However, since this inter-cell interference is received at the BS, it impacts equally all users of the given NOMA cluster. As such,  harnessing the benefits of NOMA in multi-cell scenarios requires exploitation of suitable interference mitigation strategies. For example, opportunistic NOMA may be exploited at each BS  to opportunistically decide whether to operate in NOMA or OMA mode given the interference conditions,  locations of the  users, and their vicinity from the neighboring BSs.  

\subsection{User Grouping/Scheduling, Power Allocation, and Inter-cell Interference Mitigation}
Theoretically, the number of users in a NOMA cluster can be unbounded, i.e., all users can utilize the available resources (i.e., bandwidth, time, antenna, etc.) simultaneously. However, in practice, NOMA is limited by the intra-cell/inter-cell interferences and SIC propagation errors. Therefore, to guarantee NOMA gains on a per-user basis, it is crucial to efficiently select as well as limit the number of users in a NOMA cluster. Also, after clustering, dynamic power control/allocation can be done within a cluster and/or among the clusters to achieve different performance objectives such as throughput maximization or fairness among users in different clusters.
Further, to serve the  cell-edge users, it is plausible to exploit coordination among neighboring BSs for interference avoidance.
For instance, in the downlink, cell-edge users are more vulnerable to inter-cell rather than intra-cell interference. Therefore, exploiting coordinated multi-point transmissions can mitigate interference at the cell-edge users. Also, it is possible to integrate NOMA with other multiple access schemes such as frequency hopping-based schemes for dynamic channel access. Cross-layer (PHY-MAC) design of NOMA considering adaptive modulation and coding should also be investigated to improve the  performance of the system.

\subsection{NOMA-Based Wireless Backhauling for Small Cells}

The ultra-dense nature of the future small cell deployments pushes the network operators to  investigate efficient options for simultaneous wireless backhauling of multiple small cells. In this context, various recent research studies consider optimizing the backhaul spectrum allocation, utilize massive MIMO technology to support simultaneous backhaul transmissions, and full-duplex backhaul transmissions. However, the massive MIMO technology requires computationally expensive precoding schemes at the BSs to orthogonalize the backhaul transmissions. On the other side, the full-duplex backhaul transmissions require analog/digital self-interference cancellation circuits to reduce the self-interference. In contrast, to mitigate the severe intra-cell interferences, NOMA utilizes conventional SIC methods and takes the advantage of transmit power allocations in the downlink and  distinct channel gains of users in the uplink. The application of NOMA in the backhaul transmissions of small cells is thus interesting where a given backhaul hub can schedule several small cells simultaneously.

\subsection{Cooperation in NOMA}
The use of SIC in downlink NOMA  require strong channel users to decode the  information of weak channel users perfectly. This information can be used in several ways to exploit cooperation among users of a NOMA cluster. For instance,  device-to-device (D2D) transmissions can be exploited within a NOMA cluster to broadcast/transmit the decoded message signals of  weak channel users. However, if the strong and weak channel users are significantly apart, D2D transmission may not be possible. In such a case, a relay-based cooperative communication can be exploited to enhance the diversity and performance of the cell-edge users. Further, the  appropriate mode of cooperation (i.e., relay-based cooperation or direct D2D transmission) can be selected opportunistically to minimize the latency or to enhance the throughput performance.

\section{Conclusion}
We have described the working principles and distinct features  of uplink and downlink NOMA transmissions from various perspectives such as receiver complexity, SIC, and intra-cell/inter-cell interferences. 
Then, for both  downlink and uplink NOMA,  we have derived the NOMA dominant condition  for each individual user in a two-user NOMA cluster.
The derived conditions are  distinct for uplink and downlink as well as for each individual user in a two-user NOMA cluster.
Numerical results have showed the significance of the derived conditions in selecting appropriate users for two-user uplink/downlink NOMA clusters while providing a comparison to the random user clustering.
An up-to-date literature review is then provided  to stress the existing research gaps. Finally, we have highlighted key challenges of NOMA  and future research directions.

\end{document}